\pdfoutput=1 
\documentclass[
  aps,%
  twocolumn,%
  pra,%
  groupedaddress,% 
  superscriptaddress,%
  showpacs,%
  letterpaper,%
  amsfonts,%
  floatfix%
]{revtex4-1}

\RequirePackage{amsmath,amssymb,bm,wasysym}
\RequirePackage{graphicx}
\RequirePackage{hyperref}
\hypersetup{%
  breaklinks = {true},
  citecolor = {blue},
  colorlinks = {true},
  linkcolor = {blue},
  pdfauthor = {\textcopyright\ Matthias Droth },
  pdfcreator = {\LaTeX\ and \flqq hyperref\frqq},
  pdffitwindow = {true},
  pdfmenubar = {true},
  pdfpagelayout = {OneColumn},
  pdfstartview = {FitH},
  pdftoolbar = {true},
  plainpages = {false},
  pdfpagemode = {UseThumbs},
}

\DeclareMathAlphabet{\mathpzc}{OT1}{pzc}{m}{it}
\renewcommand{\vec}[1]{\boldsymbol{#1}}
\newcommand{\teq}{{\,=\,}}
\graphicspath{ {./Figs/} } 

\begin{document}
\pagestyle{plain}
% ---------------------------------------------------------------------------- %
\title{Piezoelectricity in planar boron nitride via a geometric phase}

\author{Matthias Droth}
\thanks{Corresponding author: matthias.droth@uni-konstanz.de}
\affiliation{Department of Physics, University of Konstanz, 
78457 Konstanz, Germany}

\author{Guido Burkard}
\affiliation{Department of Physics, University of Konstanz, 
78457 Konstanz, Germany}

\author{Vitor M. Pereira}
\affiliation{%
Centre for Advanced 2D Materials \& Department of Physics, %
National University of Singapore, %
2 Science Drive 3, Singapore 117542 %
}


% ---------------------------------------------------------------------------- %
\begin{abstract}
Due to their low surface mass density, two-dimensional materials with a strong 
piezoelectric response are interesting for nanoelectromechanical systems with 
high force sensitivity. Unlike graphene, the two sublattices in a monolayer of 
hexagonal boron nitride (hBN) are occupied by different elements, which breaks 
inversion symmetry and allows for piezoelectricity. This has been confirmed with 
density functional theory calculations of the piezoelectric constant of 
hBN. Here, we formulate an entirely analytical derivation of the 
electronic contribution to the piezoelectric response in this system based on 
the concepts of strain-induced pseudomagnetic vector potential and the modern 
theory of polarization that relates the polar moment to the Berry curvature. Our 
findings agree with the symmetry restrictions expected for the hBN lattice and 
reproduce well the magnitude of the piezoelectric effect previously obtained 
\emph{ab-initio}.
\end{abstract}
\maketitle

% ---------------------------------------------------------------------------- %
\section{Introduction}

Truly two-dimensional materials became a subject of intense research with the 
experimental isolation of graphene about one decade ago \cite{Novoselov2004, 
Eda2008, CastroNeto2009, Lin2010}. In the wake of the many developments driven 
initially by research in this graphite monolayer, other two-dimensional crystals 
such as transition metal dichalcogenides, hexagonal boron nitride (hBN), 
phosphorene, and others, have gained prominence due to rich and outstanding 
electronic, magnetic, structural, and optical properties \cite{Novoselov2005(2), 
Dean2010, Radisavljevic2011, Duerloo2012}. The prospect of stacking individual 
monolayer materials with different properties holds the promise of a new 
paradigm in solid state physics as this modular concept of layered van der Waals 
heterostructures might enable the tailoring of physical properties to levels 
much beyond the bandgap engineering that is mainstream in semiconductor 
heterostructures \cite{Novoselov2012, Geim2013}. A key role in such 
heterostructures would likely fall to hBN. While many two-dimensional building 
blocks are praised for their superior intrinsic electronic properties, these are 
detrimentally sensitive to interactions with substrates, other layers, and to 
contamination \cite{Nomura2006, Chen2008, Chen2008(2)}. With a large bandgap and 
a lattice mismatch of less than 2\,\% w.r.t.~graphene, hBN has the 
potential to preserve graphene's celebrated properties within such 
heterostructures and is currently the insulating substrate of choice for clean, 
atomically flat deposition or interfacing of two-dimensional crystals 
\cite{Robertson1984, Dean2010, Slawinska2010, Britnell2012, Paszkowicz2002, 
Allen2016}. 

\begin{figure}[b]
\centering
\includegraphics[width=0.48\textwidth]{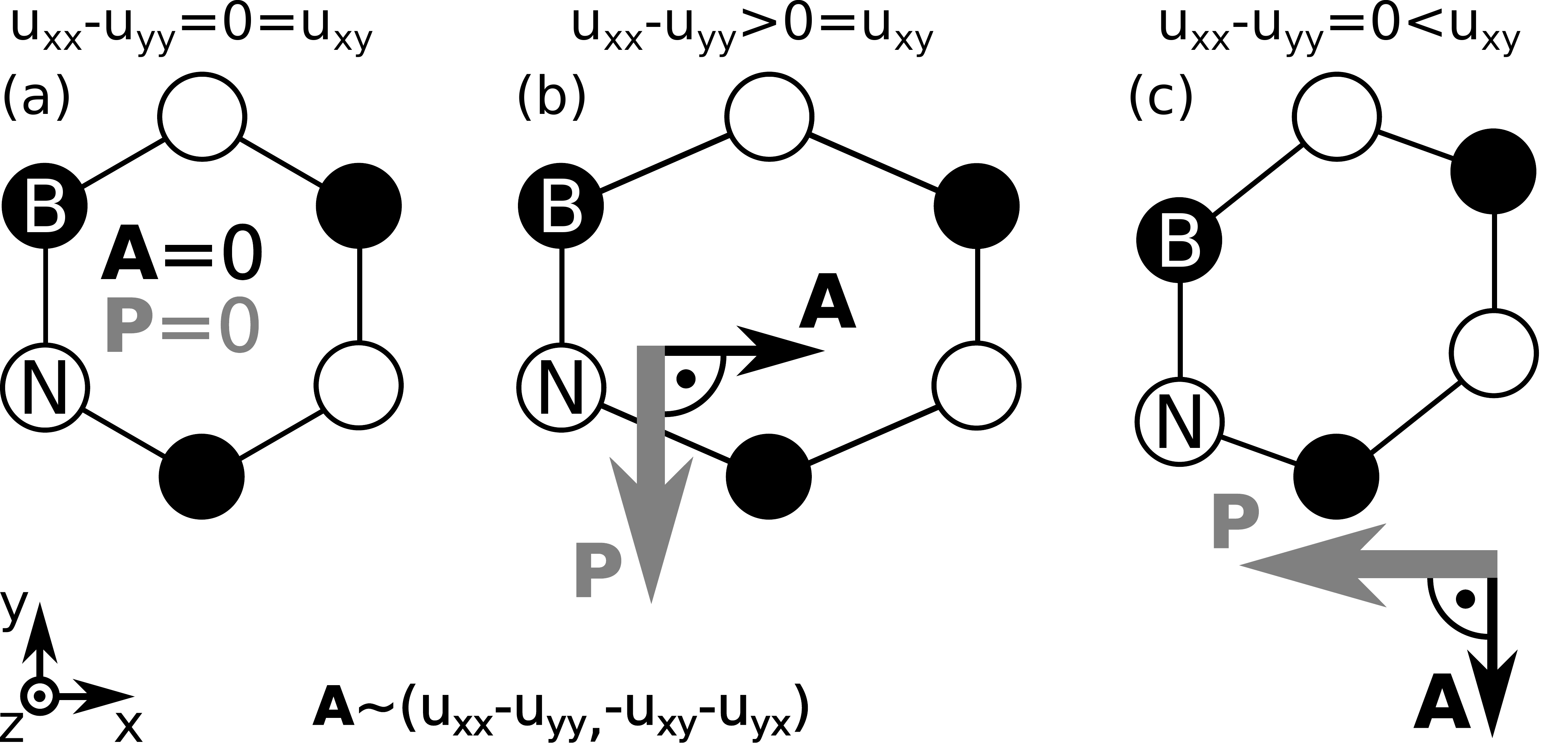} 
\caption{
The lattice of hBN does not possess an inversion center and hence 
allows for piezoelectricity. (a) Isotropic strain ($u_{xx}\teq u_{yy}, 
u_{xy}\equiv u_{yx} \teq 0$) leads to a vanishing pseudomagnetic vector 
potential 
$\vec{A}$ and does not induce a 
polarization $\vec{P}$. (b,c) Realizations of the strain tensor $u_{ij}$ that 
lift the 
trigonal symmetry and generate a change in the polarization. The 
induced polarization and the vector potential are always orthogonal, 
$\vec{P}\perp\vec{A}$.
}
\label{picBW1}
\end{figure}

Beyond such a passive role, the properties of hBN also allow for an active role. 
The monolayer of hBN has a honeycomb lattice structure similar to that of 
graphene, yet one of its two sublattices is occupied by boron (B), and the other 
by nitrogen (N) atoms, see Fig.~\ref{picBW1}. This results in a strong ionic 
bond and a bandgap of $\approx 6$\,eV \cite{Robertson1984, Watanabe2004, 
Topsakal2009,Bhowmick2011}. Since inversion symmetry is naturally absent in this 
crystal, a piezoelectric response is possible 
\cite{Mele2002,Duerloo2012,Bai2007}, i.e., a change in the bulk electric 
polarization $\vec{P}$ when subjected to external stress. 

% ----- FIGURE ----------------------------------------------------------------%
% ----- END FIGURE ------------------------------------------------------------%

The ability to control bulk polarization mechanically and, conversely, to 
convert electric fields into mechanical displacements is of enormous interest in 
the realm of energy harvesting, particularly so at the micro- and nanoscale, 
where a vision of self-powered miniaturized electronic devices is being strongly 
pursued in materials science \cite{PiezoHarvesting, Wu2014, Regan2005, Zhu2015}. On 
another front, a strong piezoelectric coupling has been shown to be an important 
tool in cooling nanoelectromechanical systems (NEMS) to their mechanical quantum 
ground state \cite{OConnell2010}. Such applications demand a strong 
piezoelectric material from the outset and that, in turn, requires a good 
insulator with a robust interplay between the underlying electronic and 
mechanic/lattice degrees of freedom. Density functional theory (DFT) 
calculations have established that an hBN monolayer has among the highest 
specific piezoelectric coefficients ($\sim 1$\,pC/N) known 
\cite{Sai2003,Duerloo2012}. Combined with the lowest surface mass density of all 
known piezoelectric crystals, this might allow for NEMS with as yet unknown 
force sensitivity \cite{Bunch2007}. Moreover, hBN could provide the 
piezoelectric coupling in a layered graphene/hBN heterostructure and thus allow 
for the electromechanical manipulation of graphene with an electric field. Its 
strong piezoelectric characteristics, high mechanical stability, and easy 
handling make hBN a prime material for such technological applications. 

The lattice of (isotropically deformed) hBN does not belong to one of the 10 
polar classes and hence, hBN exhibits no spontaneous polarization. However, it 
does sustain one when subjected to \emph{anisotropic} deformation. In this 
paper, we use the modern theory of polarization and the geometric phase 
approach \cite{Vanderbilt1990, Xiao2010, Mele2002} to calculate the 
electronic contribution to the piezoelectric tensor of hBN in an 
entirely analytical way. An ionic contribution is not considered, here 
\cite{Baroni2001, Michel2009, Michel2011}. 
A DFT calculation in the context of hBN nanotubes established that the 
dominant electronic contribution ($\approx 80\%$) to the polarization arises from the $\pi$ 
valence band, and that it has the same sign as the remaining contribution from 
the $\sigma$ valence bands \cite{Sai2003, Naumov2009, Nakhmanson2003}. 
Therefore, a minimal yet promising ansatz for the analytical description of 
piezoelectricity in hBN consists in focusing entirely on the $\pi$ bands. As we 
demonstrate, the low energy bandstructure of the $\pi$ bands is already 
sufficient to derive results in qualitative and good quantitative agreement with DFT 
calculations. 

% ---------------------------------------------------------------------------- %
\section{Model details}

Due to the underlying honeycomb lattice, low energy electrons in hBN effectively 
behave as (massive) Dirac fermions. This is similar to the situation in 
graphene, except for the mass term associated with the different orbital 
energies at the B and N atoms. The difference in electronegativity between the 
two species causes electron transfer from B to N within the $\sigma$ bonds 
\cite{Transfer} and results in a bond with an ionic character, in contrast to 
the purely covalent bond of graphene \cite{Robertson1984, Topsakal2009}. The 
broken sublattice symmetry gives rise to the bandgap \cite{Hunt2013}. Such a 
model has already been used to describe the chirality-dependent piezoelectric 
response of hBN nanotubes \cite{Mele2002}. In the vicinity of the $K$ points in 
the hexagonal reciprocal lattice, the effective Hamiltonian is then
\begin{eqnarray}
  \mathcal{H}^{(\tau)}=\hbar v_F \, \bm{\sigma}^{(\tau)}\cdot(\bm{q}-\tau
  \bm{A})
  \,,
  \label{hamilBN} 
\end{eqnarray}
where $\hbar v_F \equiv\frac{3}{2}|t|a$, with $|t|$ as the magnitude of the 
nearest neighbor tight-binding hopping amplitude and $a$ as the interatomic 
distance. 
In the absence of strain, $\bm{A}\teq 0$. In our notation, $\bm{q} \equiv 
(q_x,\,q_y,\,\Delta)$, where $q_{x,y}$ are the Cartesian components of the 
electron's crystal momentum measured relative to the high-symmetry points $K$ 
($\tau\teq {+}1$) and $K'$ ($\tau\teq {-1}$). The vector 
$\bm{\sigma}^{(\tau)}\equiv (\tau\sigma_x,\sigma_y,\sigma_z)$ is defined in 
terms of the three Pauli matrices that address the sublattice degree of freedom 
(pseudospin) in this problem. Since the real electron spin does not play a role 
in the following, it is not made explicit in our expressions. The sublattice 
potential $\hbar v_F \Delta$ arises due to the different on-site energies at the 
boron ($\Delta{\,>\,}0$) and nitrogen ($-\Delta$) atoms, and gives rise to a 
bandgap of $2\hbar v_F \Delta$ in hBN. When $\bm{A}\teq 0$, the energy 
dispersion associated with Eq.~\eqref{hamilBN} is the hyperbola 
\begin{equation}
E_{c,v}(\bm{q}) \teq \pm \hbar v_F\, (q_x^2 + q_y^2 + 
\Delta^2)^{1/2}\label{eigenEn}
\end{equation}
centered at each $K$ point. The vector $\bm{A}$ encodes the electron-lattice 
coupling for anisotropic strains and provides the essential mechanism through 
which the system can sustain a strain-induced polarization. It is well known 
that the effect of such lattice deformations can be accounted for via the 
pseudomagnetic vector potential $\vec{A}$ in the Hamiltonian 
$\mathcal{H}^{(\tau)}$. The form of Eq.~\eqref{hamilBN} aptly reflects the 
minimal-type substitution $\vec{q} \mapsto \vec{\tilde{q}} \equiv \vec{q} 
-\tau\vec{A}$, where the pseudomagnetic vector potential is given by 
\cite{KaneMele, Suzuura2002, CastroNeto2009}
\begin{eqnarray}
  \begin{pmatrix}A_x\\A_y\\A_z\end{pmatrix}
  \equiv \frac{3t\beta\kappa}{4\hbar v_F}
  \begin{pmatrix}u_{xx}-u_{yy}\\-u_{xy}-u_{yx}\\0\end{pmatrix}
  \,.
  \label{pseudoA} 
\end{eqnarray}
This definition is given in terms of the strain tensor $u_{ij} \equiv 
(\partial_iu_j + \partial_ju_i)/2$, where $\vec{u}(\vec{r})$ is the local 
displacement field, in general a function of position, although here we shall 
focus on strictly uniform and planar strain configurations. The parameter $\beta 
\equiv \frac{a}{t}\frac{\partial t}{\partial a} < 0$ describes the variation of 
the hopping amplitude $t < 0$ w.r.t.~bond length in linear order and $\kappa\sim 
1$ \cite{Suzuura2002}. In Sec.~\ref{coupling}, we discuss the complete 
prefactor/coupling strength in detail. The electron-phonon coupling encoded in 
$\vec{A}$ has the qualitative effect of displacing the center of the Fermi 
circles from $\tau\vec{K}$ to $\tau(\vec{K} + \vec{A})$ \cite{Pereira2009}. This 
affects the Berry curvature in the parameter space $(q_x,q_y,\lambda)$, where 
$\lambda{\in}[0,\Delta]$ parametrizes the sublattice potential. 

In a microscopic description of strain-induced electrical polarization, one 
formally and conventionally identifies two additive contributions 
\cite{Gironcoli1989, Resta1994, Baroni2001}. 
The first one, so called ionic contribution, 
arises from the breakdown of the Cauchy-Born rule and the need to explicitly 
consider relative ionic displacements within the crystal's unit cell that are 
not accounted for by the macroscopic strain field $u_{ij}$. 
This ionic contribution can 
be characterized analytically if an accurate empirical force constant model to 
describe the lattice degrees of freedom is known \cite{HarrisonESPS}. In the 
case of hBN, such calculations were performed by Michel and Verberck 
\cite{Michel2009, Michel2011}. %who identified an extremely high 
%value of the ionic part of the piezoelectric constant, $e_{222}{\,\simeq\,} 
%-2.43\times 10^{-10}$\,C/m, for the monolayer \cite{MichelVerbeck2011}.
%
The second, electronic contribution arises from the induced electronic 
density and is the specific focus of this paper, computed within the quantum 
phase approach \cite{Vanderbilt1990, Resta1994}. 
From an \emph{ab-initio} standpoint, a common strategy to identify these two 
contributions consists in (i) computing the \emph{electronic} polarization 
as a function of strain while keeping the ions clamped and (ii) performing the 
same computation with fully relaxed ions, which yields the \emph{total} 
(electronic and ionic) polarization. Note that the ionic and electronic 
contributions have opposite sign for hBN 
\cite{Baroni2001, Duerloo2012, Michel2011, Transfer}. 

To determine the electronic contribution to the piezoelectric constant of 
an hBN monolayer, we conceive a gedankenexperiment in which the gap 
parameter in Eq.~\eqref{hamilBN} varies adiabatically from $\lambda=0$ 
(graphene) to $\lambda=\Delta$ (hBN). Such adiabatic change is accompanied by 
the development of a bulk polarization --- induced dipole moment per unit area 
--- whose magnitude is obtained from \cite{KingVanderbilt,Mele2002,Xiao2010} 
\begin{eqnarray}
  P_i = 2\text{e}
  \sum_{\tau}\int_0^{\Delta}\text{d}\lambda\int_{\text{BZ}/2}\frac{\text{d}
  ^2q}{(2\pi)^2}\, \Omega^{(\tau)}_{q_i,\lambda}
  \,.
  \label{dpi} 
\end{eqnarray}
Here, $P_i$ is the $i$-th Cartesian component of the induced polarization 
vector, $\text{e} = |\text{e}|$ is the unit charge, and the factor of 2 accounts 
for the spin degeneracy. The integral over half the Brillouin zone (BZ) around 
each valley combined with the summation over $\tau$ recovers the full BZ 
integral required. The Berry curvature is given by
\begin{eqnarray}
  \Omega^{(\tau)}_{q_i,\lambda} \equiv 
  i\Big\langle\frac{\partial u_{\tau}}{\partial
  q_i}\Big|\frac{\partial u_{\tau}}{\partial\lambda}\Big\rangle
  -i\Big\langle\frac{\partial 
  u_{\tau}}{\partial\lambda}\Big|\frac{\partial u_{\tau}}{\partial 
  q_i}\Big\rangle
  \,,
  \label{berryc} 
\end{eqnarray}
where $|u_{\tau}\rangle$ is a valence eigenstate of Eq.~\eqref{hamilBN} with 
$\Delta \mapsto \lambda$.

% ---------------------------------------------------------------------------- %
\section{Strain-induced polarization}

From Eq.~\eqref{berryc}, one straightforwardly resolves the Berry 
curvatures
\begin{eqnarray}
  \Omega^{(\tau)}_{q_x,\lambda} = -\tau\frac{\tilde{q}_y}{2\epsilon^3}
  ,\qquad
  \Omega^{(\tau)}_{q_y,\lambda} = \tau\frac{\tilde{q}_x}{2\epsilon^3}
  \,,
  \label{berryqx} 
\end{eqnarray}
where $\epsilon \equiv |E_{c,v}(\tilde{q}_x,\tilde{q}_y,\lambda)|$. 
To be specific, we consider now the calculation of $P_x$ according to 
Eq.~\eqref{dpi}. Integration over the adiabatic parameter leads to
\begin{eqnarray}
  \int_0^{\Delta}\text{d}\lambda\frac{\tau\tilde{q}_y}{\epsilon^3} 
  =
  \frac{\tau\Delta\tilde{q}_y}
  {(\tilde{q}_x^2+\tilde{q}_y^2)\sqrt{\tilde{q}_x^2+\tilde{q}_y^2+\Delta^2}}
  \,.
  \label{DeltaInt} 
\end{eqnarray}
This is followed by the momentum integration over a square 
\mbox{$[{-}w,{+}w]^2$} centered at each of the two high symmetry points 
$K$ and 
$K'$ of the undistorted BZ. To conserve the total number of states, the 
area of 
each square is exactly half of the first BZ zone, i.e. 
$w{\,=\,}3^{3/4}K/4{\,=\,}3^{{-}3/4}\pi/a$. 
After restoring the prefactor, this BZ integration leads to an entirely 
analytical expression for $P_x$ (and likewise for $P_y$). 
For piezoelectricity, the linear response of $\vec{P}$ w.r.t.~strain is 
relevant. 
As Eq.~\eqref{pseudoA} shows that $\vec{A}$ is linear in strain, it is 
appropriate to focus on the leading order of the induced polarization in 
the pseudomagnetic vector potential:
\begin{equation}
  \vec{P} = 
    \frac{2\text{e}}{\pi^2}\tan^{-1}\bigg[\frac{\Delta}{\sqrt{2w^2{+ }\Delta^2}}
    \bigg]
    \;\vec{A}\times\hat{\textbf{z}} 
    + \mathcal{O}(A^3)
  .
  \label{PvsA}
\end{equation}
This is the main result of our analytic calculation. On one hand, it 
manifests a 
new and useful qualitative result: the pseudomagnetic vector potential 
and the 
induced  polarization are orthogonal, $\vec{P}\perp\vec{A}$, irrespective 
of the state of strain.
On the other hand, the exact and simple analytical expression in 
Eq.~\eqref{PvsA} 
allows us to extract definite quantitative predictions regarding 
the magnitude of the piezoelectric coefficient in hBN.

% ---------------------------------------------------------------------------- %
\section{Piezoelectric tensors and symmetry}

The components of the direct and converse piezoelectric tensors are, 
respectively, given by
\begin{eqnarray}
  d_{ijk} \equiv \frac{\partial P_i}{\partial\sigma_{jk}}
  \,,\qquad 
  e_{ijk} \equiv \frac{\partial P_i}{\partial u_{jk}}
  \,, 
\end{eqnarray}
where $\sigma_{jk}$ is the stress tensor.
The crystal of hBN belongs to the point group $\bar{6}m2$ ($D_{3h}$) 
which, for 
the lattice orientation introduced in Fig.~\ref{picBW1} containing a mirror 
plane perpendicular to the $x$-axis, imposes the symmetry constraints 
\begin{eqnarray}
  \begin{split}
    d_{211}=d_{112}=d_{121}=-d_{222}\,,\\
    e_{211}=e_{112}=e_{121}=-e_{222}\,,
  \end{split}
  \label{symConstraint}
\end{eqnarray}
while all other components vanish identically \cite{Nye}. 
The piezoelectric response is thus characterized by only one number, 
and we call 
$d_{222}$ and $e_{222}$ the direct and converse piezoelectric 
constants, respectively.
The direct and converse effects are related through the elastic tensor, 
$e_{ijk} \teq d_{imn} C_{jkmn}$, which, since we have only one 
independent 
component in each, reduces to the simple relation $e_{222} \teq 
d_{222}(C_{2222}{-}C_{2211})$. 
To confirm consistency of our model with symmetry constraints, note that 
Eq.~\eqref{PvsA} 
implies $(\partial P_x/\partial A_y)=-(\partial P_y/\partial A_x)$. Together 
with Eq.~\eqref{pseudoA}, 
this leads to, e.g., 
\begin{equation}
  e_{222} 
  = \frac{\partial P_y}{\partial A_x}\frac{\partial A_x}{\partial u_{yy}}
  = -\frac{\partial P_x}{\partial A_y}\frac{\partial A_y}{\partial u_{xy}}
  = -e_{112}
  \,.
  \label{e112}
\end{equation}
Analogously, we can verify that all the relations in 
Eq.~\eqref{symConstraint} 
are indeed satisfied by the model. 
The piezoelectric constant is explicitly given by
\begin{equation}
  e_{222} = \frac{\text{e}|\beta|\kappa}{\pi^2 a}
    \tan ^{-1}\biggl[\frac{\Delta}{\sqrt{2 w^2+\Delta^2}}\biggr]
  .
  \label{e222}
\end{equation}
%

% ---------------------------------------------------------------------------- %
\section{Magnitude of the piezoelectric constant}
\label{coupling}

The electromechanical coupling strength $\frac{3}{4}t\beta\kappa$ in 
Eq.~\eqref{pseudoA} arises from a low energy approximation of a tight-binding 
Hamiltonian that describes electronic hopping among the $p_z$ orbitals of 
neighboring atoms in the honeycomb lattice of hBN \cite{Suzuura2002}. Under 
strain, the interatomic distances are changed and the hopping amplitude $t < 0$ 
is modified accordingly. This couples the electronic system to the lattice 
degrees of freedom to an extent that is controlled by the parameter $\beta 
\equiv \frac{a}{t}\frac{\partial t}{\partial a} < 0$ which reflects the 
sensitivity of the hopping amplitude to changes in the bond length. Since, as in 
graphene, the $\pi$ band comes about due to electron hopping between the 
$p_z$-orbitals of nearest-neighboring atoms, $t$ corresponds to the 
Slater-Koster parameter $V_{pp\pi}$, and it is natural to expect an exponential 
decay of $t$ with increasing interatomic distance \cite{SlaterKoster,Papa}. In 
analogy with a parametrization that is fairly accurate in graphene 
\cite{Pereira2009, Ribeiro2009}, we consider
\begin{eqnarray}
  t(a) = t_0 \, e^{\beta(a - a_0)/a_0}
  \,,
  \label{tDecay}
\end{eqnarray}
where $t_0 \equiv t(a_0)$ denotes the hopping amplitude at the equilibrium bond 
length $a_0$. We can estimate $\beta$ from existing data for the Slater-Koster 
parameters $V_{pp\pi}$ in hBN with first-, second-, and third-nearest neighbors 
that are fit to accurately reproduce the bandstructure obtained from DFT 
calculations \cite{Topsakal2009, Slawinska2010, Duerloo2012, GiraudThesis}. Both 
the first- and the third-nearest neighbor hopping occur between B and N atoms. 
With $t_0 \teq t^{(1)} \teq t(1.44\,\text{\AA}) \teq {-}2.16\,\text{eV}$ and 
$t^{(3)} \teq t(2.88\,\text{\AA}) \teq {-}0.08\,\text{eV}$ for the first and 
third neighbor hopping amplitudes, respectively, we find $\beta=-3.3$, which is 
a value very similar to that for graphene \cite{Ribeiro2009}. This similarity is 
not surprising as the atomic orbitals involved are the same in the two systems, 
and the relaxed interatomic distance is nearly the same, too. For consistency of 
$t^{(1)}$ and $t^{(3)}$, we have used the values from Ref.~\cite{GiraudThesis} 
in our estimate for $\beta$. However, literature values for the nearest neighbor 
hopping amplitude cumulate rather around $t_0 \teq -2.3\,\text{eV}$ and we thus 
infer $t\beta \teq 7.6$\,eV \cite{Robertson1984, Slawinska2010, Ribeiro2011(2)}. 
The dimensionless parameter $\kappa$ depends on microscopic details of the 
lattice dynamics. In lowest order of a valence-force-field model 
\cite{Suzuura2002, HarrisonESPS}, it is given by \mbox{$\kappa\teq 1/\sqrt{2}$}. 
This results in an electromechanical coupling strength of 
$\frac{3}{4}t\beta\kappa = 4.0$\,eV in Eq.~\eqref{pseudoA}. Putting this 
together with all the other relevant parameters listed in 
Table~\ref{decayTable}, we evaluate Eq.~\eqref{e222} and finally estimate the 
contribution of the $\pi$ electrons to the total piezoelectric constants to be
\begin{eqnarray}
  \begin{split}
    e_{222} & \simeq 0.63\,\text{e/nm} = 1.0 \times 10^{-10}\,\text{C/m},\\
    d_{222} & \simeq 2.8\,\text{e/$\mu$N} = 0.44\,\text{pC/N}.
  \end{split}
  \label{estimate}
\end{eqnarray}
Duerloo \emph{et al.} obtained the values $e_{222} \simeq 1.38 \times 
10^{-10}$\,C/m in a fully relaxed-ion DFT calculation and $e_{222} \simeq 
3.71 \times 10^{-10}$\,C/m under clamped-ion conditions (note that our 
coordinate convention is different from the one used by these 
authors) \cite{Duerloo2012}. As described earlier, the latter corresponds to 
the electronic contribution and is the one appropriate for direct comparison 
with the figures quoted in Eq.~\eqref{estimate} since our model accounts 
only for the electronic part. 
We further recall that our calculation hinges entirely on the electronic 
effects associated with $\pi$-band electrons, justified by the fact that, 
according to first principles calculations, these account for 80\,\% of the 
electronic piezoelectric response \cite{Sai2003}. This factor of $0.8^{-1}$ can 
be incorporated in our results allowing us to refine the numbers in 
Eq.~\eqref{estimate} to $e_{222} \simeq 1.3 \times 10^{-10}$\,C/m and $d_{222} 
\simeq 0.55$\,pC/N as the prediction for the electronic contribution to the 
piezoelectric constants.
Given the uncertainty inherent to our estimate of the electromechanical coupling 
strength $\frac{3}{4}t\beta\kappa$ above, we consider this result to be in good 
quantitative agreement 
with the first principles value of $3.71 \times 10^{-10}$\,C/m. 
Beyond the scope of the current work, it would be interesting to 
obtain \emph{ab-initio} a more precise value of the logarithmic derivative of 
the hopping, $\beta$, so the quantitative accuracy of our model can be 
fully assessed.

We reiterate the impressive magnitude of the piezoelectric response in hBN 
already pointed out by Duerloo \emph{et al.} as well as Michel and Verberck 
\cite{Michel2011}. For comparison, the piezoelectric tensor components of 
quartz vary in the range $d \sim 0.7-2.3$\,pC/N \cite{Bechmann1958}. The 
numbers in Eq.~\eqref{estimate} show that a single, atomically thin layer of 
hBN is essentially as good a piezoelectric as a crystal of quartz.

% ----- TABLE -----------------------------------------------------------------%
\begin{table}
\centering
\begin{tabular}{ll|ll}
\hline\hline
Quantity & Value & Quantity & Value\\
\hline
$\beta$&$-3.3$  &  $\kappa$&$0.71$\\
$w$&$0.96\,\text{\AA}^{-1}$  &  $a$&$1.44$\,\AA\\
$\Delta$&$0.60\,\text{\AA}^{-1}$ & 
$C_{2222}-C_{2211}$&$229\,\text{N\,m}^{-1}$\\ 
\hline\hline
\end{tabular}
\caption{
The parameter $\beta$ is discussed around Eq.~\eqref{tDecay} and $\kappa$ 
follows from a lowest order valence-force-field model \cite{Suzuura2002, 
HarrisonESPS}. The value for $a$ stems from Refs.~\cite{Slawinska2010, 
Paszkowicz2002} and is also used for $w = 3^{-3/4}\pi/a$. With $t = 
-2.3$\,eV \cite{Robertson1984, Slawinska2010, Ribeiro2011(2)} and $\hbar v_F = 
3|t|a/2$, $\Delta$ corresponds to a bandgap of $2\hbar v_F\Delta=6.0$\,eV 
\cite{Watanabe2004, Slawinska2010, Bhowmick2011, Topsakal2009}. For 
$C_{2222}-C_{2211}$, we use the elastic constants $C_{11} \teq 291$\,N/m and 
$C_{12}\teq 62$\,N/m, as reported in Ref.~\cite{Duerloo2012} (Voigt notation; 
different lattice orientation).}
\label{decayTable} 
\end{table}
% ----- END TABLE -------------------------------------------------------------%

% ---------------------------------------------------------------------------- %
\section{Conclusion}

We have obtained exact analytical results for the induced polarization and 
piezoelectric constant of monolayer hBN within the quantum geometric phase 
approach. In our minimal model, which is proven to satisfy the symmetry 
constraints expected for the point group $\bar{6}m2$ ($D_{3h}$), the interaction 
between deformations and the electronic degrees of freedom is captured in the 
effective two-band Hamiltonian via a pseudomagnetic vector potential. 
Ionic degrees of freedom are not considered. Using 
existing literature estimates for the relevant bandstructure parameters and 
elastic constants in this system, we find that the converse and direct 
piezoelectric constants for this model are as high as $e_{222} \teq 1.3 \times 
10^{-10}$\,C/m and $d_{222} \teq 0.55$\,pC/N, respectively. The strain-induced 
polarization $\vec{P}$ is exactly perpendicular to the pseudomagnetic vector 
potential $\vec{A}$.

We also provide an estimate for the so far unknown coupling strength of the 
strain-induced pseudomagnetic vector potential in hBN, namely, 
$\frac{3}{4}t\beta\kappa \teq 4.0$\,eV. That the magnitude of the piezoelectric 
coefficient obtained here agrees well with the value extracted from independent 
DFT calculations attests to the validity and pertinence of the minimal model, 
especially since it provides a simple analytical result for its dependence on 
the basic material parameters. Another advantage of our calculation is that, 
through Eqs.~\eqref{hamilBN} and \eqref{PvsA}, one can interpret the 
piezoelectric effect in this material as a consequence of the displacement of 
the Dirac point under strain from its default position at $K$ in the BZ: a 
measurement of the electric polarization is thus an indirect measure of how 
much and along which direction the Dirac point drifts from $K$ under strain.

Our findings for atomically flat hBN might be ultimately put to test in 
hopefully upcoming experiments which, to our knowledge, have not been 
reported yet. We thus provide an analytical and concise description of 
piezoelectricity in hBN that is of relevance for the understanding of nanoscale 
devices containing hBN as a piezoelectric component, including \mbox{--- but not 
limited to --- }\newpage 
\noindent heterostructured NEMS based on two-dimensional materials. 

% ---------------------------------------------------------------------------- %
\section{Acknowledgements}

MD and GB thank the European Science Foundation and the Deutsche 
Forschungsgemeinschaft (DFG) for support within the EuroGRAPHENE project CONGRAN 
and the DFG for funding within SFB 767 and FOR 912. 
VMP was supported by the National Research Foundation (Singapore) under its Medium Sized Centre Programme and CRP grant ``Novel 2D materials with tailored properties: Beyond graphene'' (Grant No. NRF-CRP6-2010-05).

% ---------------------------------------------------------------------------- %


\begin{thebibliography}{}
\bibitem{Transfer}
The transfer of negative charge from B to N within the $\sigma$ bond is 
overcompensated by the charge transfer within the $\pi$ bond which 
gives the 
boron atom an effective negative charge in equilibrium 
\cite{HarrisonESPS}.
\end{thebibliography}
\end{document}